\documentclass[aip, reprint]{revtex4-1}

\usepackage{graphicx}
\usepackage{dcolumn}
\usepackage{bm}
\usepackage[utf8]{inputenc}
\usepackage[T1]{fontenc}
\usepackage{mathptmx}
\usepackage{etoolbox}
\usepackage{braket}
\usepackage{amsmath}

\makeatletter
\def\@email#1#2{%
 \endgroup
 \patchcmd{\titleblock@produce}
  {\frontmatter@RRAPformat}
  {\frontmatter@RRAPformat{\produce@RRAP{*#1\href{mailto:#2}{#2}}}\frontmatter@RRAPformat}
  {}{}
}%
\makeatother
\begin{document}

\title{DRAG-Inspired Single Flux Quantum Control of Superconducting Qubits}

\author{V. Vozhakov}
\email{sevozh@yandex.ru}
\affiliation{Joint Venture Quantum Technology}

\author{M. Bastrakova}
\affiliation{Joint Venture Quantum Technology}
\affiliation{Lobachevsky State University of Nizhny Novgorod}

\author{A. Liniov}
\affiliation{Lobachevsky State University of Nizhny Novgorod}

\author{D. Kulandin}
\affiliation{Independent Researcher}

\author{N. Klenov}
\affiliation{Moscow Technical University of Communications and Informatics (MTUCI)}

\author{I. Soloviev}
\affiliation{Joint Venture Quantum Technology}
\affiliation{Skobeltsyn Institute of Nuclear Physics, Lomonosov Moscow State University}

\date{\today} 

\begin{abstract}
Scalable control remains a major challenge for superconducting quantum processors due to the wiring, thermal load, and hardware complexity associated with conventional microwave-based architectures. A framework for qubit control based on Single Flux Quantum (SFQ) digital electronics is presented. For transmon qubits, optimized control sequences are constructed from regular pulse trains framed by arbitrary subsequences (ramps), enabling DRAG-inspired leakage suppression in the discrete SFQ regime through genetic-algorithm optimization. For representative $Y_{\pi/2}$ gates, the resulting sequences achieve simulated coherent-control infidelities below $10^{-4}$, including leakage from the computational subspace. The objective of the proposed approach is to demonstrate that coherent errors intrinsic to SFQ control can be reduced to a comparable level for a more scalable cryogenic control architecture.

Memory requirements are reduced by storing only the optimized ramps, while a modular controller architecture supporting both regular and arbitrary pulse generation is proposed. The logical complexity of this architecture is estimated from the Josephson-junction counts of its constituent SFQ cells. The proposed approach therefore combines low coherent-control error with reduced memory overhead and provides a basis for scalable cryogenic SFQ control architectures.
\end{abstract}

\maketitle

\section{Introduction}\label{sec:introduction}
Superconducting qubits are among the most promising physical candidates for a large-scale, fault-tolerant quantum computer. This technological platform offers long coherence times \cite{coherence, coherence2} and high-fidelity multi-qubit gates \citep{goodgates, goodgates2}. One of the major challenges in superconducting quantum computing is scaling systems to thousands of qubits. A major obstacle arises from the limitations of microwave-based control architectures, specifically the unwanted crosstalk and thermal noise introduced by hardware into the cryostat.

Furthermore, the amount of specialized hardware required for managing large-scale qubit systems adds substantially to the complexity, as conventional control architectures require a large number of microwave-generation, waveform-control, routing, and readout channels whose resource requirements grow with the number
of qubits \citep{krantz}. Some partial remedies involve utilizing custom FPGAs (Field Programmable Gate Arrays) \citep{fpga} and reducing the number of independent microwave sources for intermediate-scale qubit processors; however, these issues remain unresolved for future processors intended to handle thousands of qubits. Various approaches have been proposed to tackle this problem. Among the proposed approaches are cryo-CMOS \citep{cryocmos}, photonic links \citep{photoniclink} and superconducting digital electronics based on Single Flux Quantum (SFQ) logic. Key benefits of this approach include power efficiency \citep{sfq1, sfq2} and well-established solutions for pulse generation \citep{likharev}.

It is important to distinguish the motivation for SFQ-based control from the pursuit of the highest possible single-qubit gate fidelity. Conventional microwave control of transmon qubits is a mature technology, and experimentally measured average single-qubit errors below $10^{-4}$ have already been demonstrated \citep{Li2023}. More recent high-coherence transmon devices have reported single-qubit fidelities approaching $99.994\%$ \citep{Bland2025}. The principal motivation for SFQ control is therefore not an improvement over microwave control in intrinsic single-qubit fidelity, but the possibility of moving pulse generation, sequencing and routing into the cryogenic environment and thereby reducing the room-temperature hardware and wiring overhead associated with large-scale systems.

Experimental SFQ control nevertheless introduces its own implementation-specific limitations. In particular, quasiparticle generation associated with dissipative SFQ circuitry limited the fidelity in the on-chip implementation of Leonard \textit{et al.} \citep{dcsfq1}. The multichip architecture of Liu \textit{et al.} \citep{dcsfq2} spatially separated the SFQ circuitry from the qubit chip in order to suppress phonon-mediated quasiparticle poisoning, although an additional incoherent contribution attributed to photon-mediated quasiparticles remained. Consequently, suppression of quasiparticle generation is an important requirement for reaching the coherent-error levels considered in the present work experimentally.

A number of theoretical and experimental studies have demonstrated the feasibility of SFQ-based qubit control~\citep{dcsfq1, dcsfq2, mcdermott2014, controller_design1, controller_design2, controller_design3, controller_design4, controller_design5, controller_design6, controller_design7, controller_design8, controller_design9, controller_design10}. Experimental SFQ control has not yet reached the fidelities obtained using state-of-the-art microwave control. For example, Liu \textit{et al.} reported an error per Clifford of $1.2(1)\%$ using an SFQ multichip-module architecture~\cite{dcsfq2}, with the performance limited in part by implementation-dependent incoherent mechanisms such as quasiparticle poisoning. More recently, integrated superconducting digital control has demonstrated single-qubit fidelities above $99\%$ and up to $99.9\%$, together with digital demultiplexing of control signals among several qubits \cite{controller_design9}. These results illustrate both the remaining experimental challenges of SFQ control and its potential advantage for scalable cryogenic integration.

In this work, we address a complementary question: whether coherent errors intrinsic to SFQ control of a weakly anharmonic transmon, particularly leakage outside the computational subspace, can be suppressed below the level at which they constitute a fundamental obstacle to such a scalable architecture.

This concept can also be applied to the SFQ control framework by employing search algorithms that determine the optimal sequence for a qubit with specified parameters. The search procedure is quite intuitive because SFQ sequences are represented as binary codes, naturally discretizing the problem. One effective strategy is the SCALLOPS (SCALable Leakage Optimized Pulse Sequences) approach, which employs pulse-generation frequencies substantially higher than the qubit frequency, thereby reducing the overall sequence duration \citep{sfq_leakage}. Our prior research has shown that, by modifying the control scheme and incorporating pulses of varying polarities, it is possible to further reduce the sequence length while maintaining essentially the same level of unwanted leakage \citep{bipolar}.

Another problem that arises with this approach is memory usage: the entire sequence must be pre-stored before it can be sent to the qubit. Previously, we have demonstrated that at higher generator frequencies, the control sequence can comprise several hundred pulses \citep{bipolar}. Moreover, there exists a trade-off between gate parameters and memory requirements: increasing the pulse generation frequency expands the optimal control landscape, potentially enabling sequences that correspond to quantum gates with higher fidelities and shorter durations. At the same time, higher frequencies imply greater flexibility in pulse placement within a single qubit cycle, consequently requiring increased memory capacity.

One potential resolution is to seek control sequences composed of repetitive subsequences, as demonstrated in \citet{sfq_leakage}. However, this strategy may be difficult to implement, because enforcing constraints on the target sequence structure makes the search process more complex. This is particularly relevant in light of works suggesting that quantum control landscapes may exhibit an arbitrarily large number of local minima, which are often difficult to locate when additional conditions are imposed on the internal structure of the control sequence \citep{notraps}.

This paper develops a methodology for optimizing SFQ-based single-qubit control sequences with particular emphasis on coherent control errors, leakage suppression, and controller-memory requirements. First, a model describing coherent transmon evolution under arbitrary SFQ pulse sequences is introduced. Second, a combined-sequence approach is investigated, in which a regular pulse train is framed by shorter arbitrary subsequences (ramps), building on the concept proposed in \citet{shillito}. A genetic algorithm-based optimization procedure is then used to identify sequences implementing a target $Y_{\pi/2}$ operation while suppressing leakage from the computational subspace. Finally, a controller architecture of the proposed control scheme is considered, including SFQ-based controllers capable of generating both regular and arbitrary pulse sequences.

The fidelities calculated in this work quantify the intrinsic coherent-control performance of the SFQ sequences. Relaxation, dephasing, quasiparticle poisoning, and other implementation-dependent incoherent mechanisms are not included in the optimization and must be considered separately when predicting the total fidelity of an experimental implementation.

\section{Theoretical framework}\label{sec:theory}

We model the transmon as a driven Kerr anharmonic oscillator with fundamental transition frequency $\omega_{01}$ and anharmonicity $\alpha$. Under the control field $\varepsilon(t)$, the Hamiltonian is:

\begin{equation}
    \hat{H}_q = \omega_{01} \hat{a}^{\dagger} \hat{a} - \frac{\alpha}{2} \hat{a}^{\dagger} \hat{a} \left( \hat{a}^{\dagger} \hat{a} - \hat I \right) - i \varepsilon \left( t \right) \left( \hat{a}^{\dagger} - \hat{a} \right).
\label{eq:transmon_hamiltonian}
\end{equation}

Throughout this work, $\hbar=1$, $\hat{I}$ denotes the identity operator. Eq.~\eqref{eq:transmon_hamiltonian} corresponds to the standard weakly anharmonic Kerr-oscillator approximation to the transmon. The qubit eigenstates $\ket{j}$ are defined as the eigenstates of the undriven Hamiltonian, $\varepsilon(t)=0$. For the main calculations, the Hamiltonian is truncated to the three lowest levels, with $\ket{0}$ and $\ket{1}$ forming the computational subspace and $\ket{2}$ representing the lowest leakage level. 

The evolution of an initial state $\ket{\psi_0}$ under the SFQ control sequence is calculated by solving the time-dependent Schr\"{o}dinger equation. The duration $\tau$ of an individual SFQ pulse is assumed to be much shorter than the qubit oscillation period. For the parameters used below, $\tau=4~\mathrm{ps}$, whereas $2\pi/\omega_{01} = 200~\mathrm{ps}$ for a $5~\mathrm{GHz}$ qubit. Thus, an individual SFQ pulse can be regarded as a short discrete rotation on the timescale of the qubit dynamics. The numerical propagation nevertheless retains the finite rectangular pulse duration; no zero-duration pulse approximation is made. The complete gate sequence contains many such pulses and extends over many qubit periods.

In terms of evolution operators:
\begin{equation}
\ket{\psi \left( t \right)} = \hat{U} \left( t \right) \ket{\psi_0}, \quad
\hat{U} \left( t \right) = \hat{P} \: e^{-i \int_0^t \hat{H}_q \left( \tau \right) d\tau}.
\label{eq:time_evolution}
\end{equation} 

Here, $\hat{P}$ denotes the time-ordering operator. The population of the $j$-th transmon level, where $j=0,1,\ldots,K-1$, $K$ is total number of levels, is defined as:
\begin{equation}
W_j(t)=\left|\braket{j|\psi(t)}\right|^2.
\label{eq:level_population}
\end{equation}

In order to quantify how accurately the sequence operator $\hat U$ implements the target single-qubit gate $\hat U_g$, we first define the projector onto the computational subspace,
\begin{equation}
\hat{\Pi} = \ket{0}\bra{0} + \ket{1}\bra{1},
\label{eq:computational_projector}
\end{equation}

and the projected evolution operator
\begin{equation}
\hat G = \hat{\Pi}\:\hat U\:\hat{\Pi}.
\label{eq:projected_evolution}
\end{equation}

Because population may leave the computational subspace, $\hat G$ is in general non-unitary. For an initial state $\ket{\psi_0}$ in the computational subspace,
the fidelity with respect to the target operation is
\begin{equation}
F_{\psi_0}
= \left| \bra{\psi_0} \hat U_g^\dagger \hat G \ket{\psi_0} \right|^2.
\label{eq:state_fidelity}
\end{equation}

No postselection or renormalization of the projected state is performed. Consequently, population transferred outside the computational subspace directly reduces the state fidelity.

The average gate fidelity is evaluated over the six cardinal states of the Bloch sphere~\cite{fidelity},
\begin{equation}
\braket{F} = \frac{1}{6} \sum_{\ket{\psi_0} \in \mathrm{S}}
F_{\psi_0}, \quad \mathrm{S}=\{\ket{x^\pm},\ket{y^\pm},\ket{z^\pm}\}
\label{eq:fidelity}
\end{equation}

where
\begin{align}
\ket{x^\pm} &= \frac{\ket{0}\pm\ket{1}}{\sqrt{2}}, \label{eq:cardinal_x}
\\
\ket{y^\pm} &= \frac{\ket{0}\pm i\ket{1}}{\sqrt{2}}, \label{eq:cardinal_y}
\\
\ket{z^+} &= \ket{0}, \qquad \ket{z^-} = \ket{1}. \label{eq:cardinal_z}
\end{align}

For a two-dimensional computational subspace, the same average can be written as
\begin{equation}
\braket{F} = \frac{ \operatorname{Tr} \left( \hat G^\dagger\hat G \right) + \left| \operatorname{Tr} \left( \hat U_g^\dagger\hat G \right) \right|^2}{6}.
\label{eq:fidelity_projected}
\end{equation}

The corresponding average leakage is
\begin{equation}
L = 1 - \frac{1}{2} \operatorname{Tr} \left( \hat G^\dagger\hat G \right).
\label{eq:leakage}
\end{equation}

Therefore,
\begin{equation}
\operatorname{Tr} \left( \hat G^\dagger\hat G \right) = 2(1-L),
\label{eq:trace_leakage_relation}
\end{equation}
and leakage contributes directly to the reported infidelity $1-\braket{F}$.

The fidelity defined above characterizes coherent control errors, including population leakage from the computational subspace. The simulations in the present work use unitary evolution and do not include environmental decoherence. Consequently, the reported infidelity should be interpreted as a coherent-control contribution to the total experimental gate error rather than as a prediction of the experimentally measurable gate fidelity. The present framework, however, can also be extended to open-system dynamics: for a specific experimental device, relaxation and dephasing can be included by replacing the unitary evolution with a Lindblad master equation.

In the weak-dissipation limit, $t_{\mathrm{gate}}\ll T_1,T_2$, ($T_1$ -- relaxation time, $T_2$ -- transverse coherence time) the corresponding first-order decoherence contribution to the average single-qubit gate infidelity is independent of the particular ideal gate \cite{Abad2022} and can be written as

\begin{equation}
r_{\mathrm{dec}} \simeq \frac{t_{\mathrm{gate}}}{6T_1}
+ \frac{t_{\mathrm{gate}}}{3T_2}.
\label{eq:decoherence_estimate}
\end{equation}

Thus, for gate durations of order $10~\mathrm{ns}$ and $T_1, T_2$ of order $100~\mu\mathrm{s}$, the decoherence contribution is expected to be of order several $10^{-5}$ and can therefore be comparable to the coherent-control infidelities reported here. For longer gates or shorter coherence times, this contribution can approach or exceed $10^{-4}$.

Only qubit control via SFQ pulses is considered. The pulses are modeled as rectangular voltage pulses with picosecond-scale duration and are assumed to be generated by a device operating at a fixed frequency $\omega_g$. Each individual SFQ pulse induces a small rotation of the qubit state. We denote the rotation angle associated with a single pulse by $\Theta$. The complete control sequence consists of many such pulses, with their timing determining the corresponding rotation axes and accumulated evolution, and is optimized to implement the desired target gate. The single-pulse rotation angle is determined by the qubit and generator parameters~\citep{mcdermott2014}:

\begin{equation}
\Theta=C_c \int_0^\tau \varepsilon_{pulse} \left( t \right) dt \: \sqrt{\frac{2\omega_{01}}{\hbar C_q}}.
\label{eq:single_pulse_rotation}
\end{equation}

It should be emphasized that $\Theta$ denotes the rotation induced by an individual SFQ pulse and not the rotation angle of the complete gate. $C_c$ and $C_q$ denote the coupling and qubit capacitances, respectively, $\tau$ is the pulse duration, and $\varepsilon_{\mathrm{pulse}}$ describes the drive field associated with a single pulse. For SFQ pulses, $\int \varepsilon_{\mathrm{pulse}}(t) dt=\Phi_0$. The calculations assume transmon parameters of $C_c = 1 \:\mathrm{fF}$ and $C_q = 1 \:\mathrm{pF}$. The following sections consider three SFQ control schemes.

\begin{description}
   \item [Regular unipolar SFQ sequence] \hfill \\ In this case, the qubit is controlled by a sequence of $m$ equidistant picosecond pulses, each with the duration $\tau$, voltage amplitude $V_0$ and generated at a frequency $\omega_g$:
   \begin{equation}
        \varepsilon \left( t \right) = V_0 \sum_{i = 1}^{m} \left[ \theta \left( t - iT \right) - \theta \left( t - iT - \tau \right) \right],
    \label{eq:regular_unipolar}
    \end{equation}
    
    Here, $\theta(t)$ is the Heaviside function, and $T = 2\pi/\omega_g$ is the pulse repetition period.
        
    \item [Regular bipolar SFQ sequence] \hfill \\ In this case, two generators are used, producing two regular pulse sequences of opposite polarity at the same frequency, with one sequence shifted by $\pi$ relative to the other:
    
    \begin{equation}
        \varepsilon \left( t \right) = V_0 \sum_{i = 1}^{m} (-1)^{i+1}\left[ \theta \left( t - iT/2 \right) - \theta \left( t - iT/2 - \tau \right) \right],
    \label{eq:regular_bipolar}
    \end{equation}
    
    \item [Optimized SFQ sequence] \hfill \\ Using optimization protocols, it is possible to determine the optimal pulse function $\varepsilon(t)$ for arbitrary qubit and generator parameters. In the SFQ case, the pulse train can be represented as a binary or ternary array of length $M$, where each generator period corresponds to one element of the array, and the array length equals the control sequence length.
    
    \begin{equation}
        \varepsilon(t) =  V_0\sum_{i = 1}^m S_i \left[ \theta \left( t - iT \right) - \theta \left( t - iT - \tau \right) \right].
    \label{eq:optimized_sfq_sequence}
    \end{equation}
    
    Here, the discrete function $S$ is a vector with length $m$ that defines the sequence pattern. In the unipolar case, the sequence is binary, $S_i \in \{0,1\}$, where “0” and “1” indicate either the absence or presence of a voltage pulse during the generator cycle, respectively. In the bipolar case, the sequence is ternary, $S_i \in \{0,1,-1\}$, with “-1“ representing the presence of a negative pulse during the generator cycle. 
\end{description}

The optimization process is a procedure aimed at finding a control operator $\hat{U}$ that maximizes the fidelity $\braket{F}$ of the realized quantum gate. In the SFQ case, it comes down to finding the optimal pulse pattern function $S$ using a discrete optimization method. This approach achieves the best results in terms of both qubit operation accuracy and speed. Previous closed-system simulations have demonstrated coherent-control infidelities of order $10^{-4}$ for SFQ sequences only several nanoseconds in length~\citep{bipolar}. However, effectively implementing this control scheme requires a significant amount of extra hardware, as the sequences can contain several hundred bits or trits that must be stored in memory cells. Currently, there are solutions for SFQ-based memory in the form of Non-Destructive Read Out (NDRO) registers, which can be used as random-access memory modules \citep{ndro_ram} and integrated into SFQ-based qubit controller architectures \citep{quantum-classical}. Additionally, there is currently no established solution for trit-level data storage, and if bipolar control schemes are employed, an additional trit-to-bit decoder would be necessary. The most common approach used in ternary computational systems is to encode a trit using two bits of information \citep{ternary}.

The numerical results presented below are obtained using the three-level Kerr model introduced above. To assess the sensitivity of the calculated gate error to this truncation, we additionally reevaluated an SFQ sequence using larger Hilbert-space truncations while keeping the pulse sequence and all physical control parameters fixed. The three-level approximation slightly underestimates the absolute coherent-control error for this sequence, while the multilevel result remains below the $10^{-4}$ criterion used throughout this work. The optimization algorithm itself is not specific to the three-level model: it operates on the discrete pulse sequence, while the fitness is obtained from numerical propagation under the chosen Hamiltonian. The same procedure can therefore be applied with a larger Hilbert-space truncation or with a more complete transmon Hamiltonian by replacing the dynamical model used in the fitness evaluation.

\section{DRAG-inspired SFQ control framework}\label{sec:drag_framework}

Fast control of a weakly anharmonic transmon can lead to population transfer outside the computational subspace, primarily through the $\ket{1} \rightarrow \ket{2}$ transition, as well as to coherent phase
errors. In conventional microwave control, these effects can be suppressed using the DRAG (Derivative Removal by Adiabatic Gate) technique~\citep{drag, drag2, drag3}. In its simplest form, DRAG  supplements the primary drive envelope with an additional quadrature proportional to its time derivative. For a primary quadrature $\Omega_I(t)$, the first-order correction can be written schematically as

\begin{equation}
    \Omega_Q(t) = -\frac{\lambda}{\alpha} \frac{d\Omega_I(t)}{dt},
\label{eq:drag_quadrature}
\end{equation}

where $\alpha$ is the qubit anharmonicity and $\lambda$ is a dimensionless scaling parameter. The resulting modification is concentrated primarily near the turn-on and turn-off regions of the pulse and provides additional control over leakage and phase accumulation.

A direct implementation of this continuous DRAG correction is not  possible with an SFQ pulse train, since an SFQ control signal consists of discrete pulses with fixed pulse area rather than continuously adjustable quadratures. Nevertheless, Shillito \textit{et al.}~\citep{shillito} showed that the structure of a DRAG-shaped control pulse can motivate an analogous organization of an SFQ sequence. In this construction, a regular pulse train responsible for the main qubit rotation is preceded and followed by shorter pulse sequences, referred to as ramps (and we will retain this terminology throughout the article). The additional timing degrees of freedom within these ramps can be used to reduce leakage and coherent phase errors associated with the beginning and end of the control sequence.

The present work adopts this \emph{structural} idea but does not attempt to reproduce the analytic DRAG correction itself. In particular, the ramps are not obtained by sampling a continuous DRAG envelope or its derivative. Instead, they are treated as discrete pulse patterns whose elements are determined directly by numerical optimization of the leakage-inclusive gate fidelity. 

The connection to DRAG should therefore be understood as a DRAG-inspired control strategy rather than as a discrete implementation of the DRAG waveform. The regular pulse train performs the principal qubit rotation, while the boundary sequences provide additional degrees of freedom for correcting the coherent errors generated by the complete driven dynamics. Suppression of spectral weight near the $\ket{1}\rightarrow\ket{2}$ transition provides a useful physical interpretation of successful ramp sequences, but it is not imposed as an optimization constraint directly. The genetic algorithm optimizes the complete gate fidelity instead.

This distinction also motivates comparing DRAG-inspired constrained ramps with ramps optimized without such restrictions. The combined sequence construction and these two optimization strategies are introduced below.

\subsection{Combined sequence approach}\label{subsec:combined_sequence}

\begin{figure}[htbp]
    \centering
    \includegraphics[width=1\linewidth]{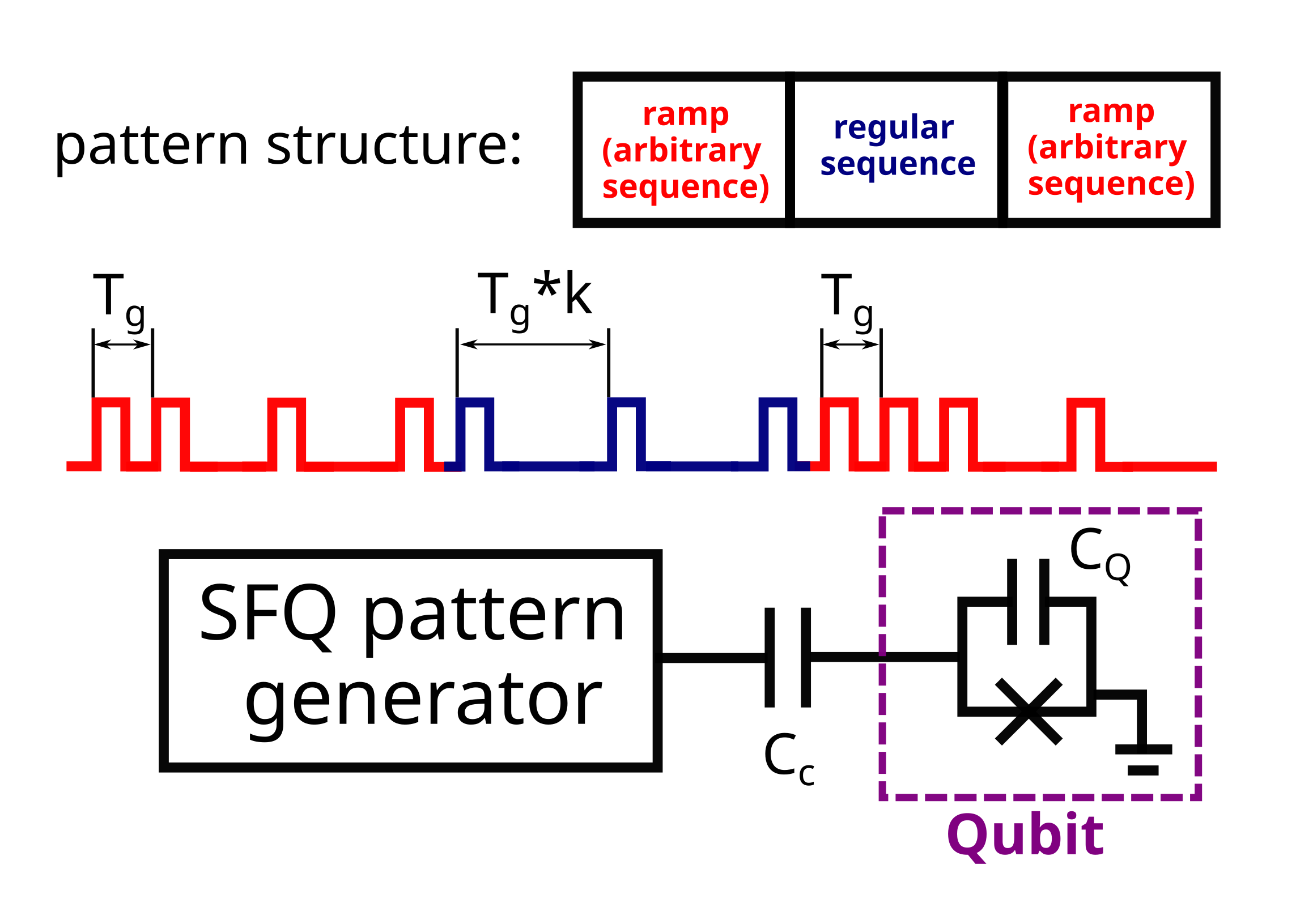}
    \caption{
    Schematic of the DRAG-inspired ramp-based SFQ control sequence. A regular sequence of picosecond-duration SFQ pulses is framed by two shorter correction sequences, referred to as the on- and off-ramps. The ramps are generated on a timing grid with a clock frequency $k$ times higher than that of the regular sequence. In the unrestricted optimization considered below, the detailed ramp patterns are determined numerically rather than obtained from a direct discretization of a continuous DRAG waveform.
    }
    \label{fig:1}
\end{figure}

Following the DRAG-inspired structure introduced above, we decompose the complete SFQ control sequence into three parts: a regular pulse train responsible for the main qubit rotation, preceded by a turn-on ramp and followed by a turn-off ramp. The ramps are generated on a finer timing grid than the regular part and provide additional degrees of freedom for suppressing leakage and coherent phase errors. The general structure of the control sequence is illustrated in Fig.~\ref{fig:1}.

The corresponding evolution operator is written as
\begin{equation}
\hat{U} = \hat{U}_{\mathrm{off}} \:
\hat{U}_{\mathrm{train}} \:
\hat{U}_{\mathrm{on}},
\label{eq:combined_evolution}
\end{equation}

where $\hat{U}_{\mathrm{train}}$ describes the evolution generated by the regular pulse train, while $\hat{U}_{\mathrm{on}}$ and $\hat{U}_{\mathrm{off}}$ are generated by the two boundary ramps.

The DRAG-inspired construction of \citet{shillito} suggests a particular relation between the turn-on and turn-off corrections. However, this relation originates from the continuous analytic DRAG construction and need not be optimal when the control space is restricted to fixed-area SFQ pulses on a discrete timing grid. We therefore use the DRAG-inspired symmetry as one reference case, but also allow the two ramps to be optimized independently.

To examine the role of this constraint, three control scenarios are considered: (i) a regular sequence without ramps, (ii) a combined sequence with DRAG-inspired symmetric ramps, and (iii) a combined sequence with independently optimized ramps and no imposed symmetry. This comparison allows us to distinguish the benefit of introducing boundary corrections from the additional benefit of removing the analytic symmetry constraint.

Throughout the optimization results presented below, the target single-qubit operation is $\hat{U}_{g}=Y_{\pi/2}$. The parameter $\Theta$ used below denotes the rotation induced by a single SFQ pulse and should not be confused with the total $\pi/2$ rotation of the target gate. The complete $Y_{\pi/2}$ operation is produced by the full optimized SFQ sequence.

We consider a transmon with following parameters, typical for qubits of this type: $\omega_{01}/2\pi = 5 \: GHz, \alpha/2\pi = 0.25 \: GHz$ and a generator producing SFQ pulses at a frequency $\omega_g = 8\omega_{01}$ with pulse width $\tau = 4$~ps, each pulse inducing a rotation by an angle $\Theta = 0.030$. This choice of parameters is convenient, as each pulse placement within the qubit period can be easily interpreted as a basic qubit operation $R_i(\phi)$, where $i$ is the rotation axis, and $\phi$ is the rotation angle:
\begin{equation}
    \begin{aligned}
        &10000000 \longrightarrow R_Y \left( \Theta \right), \\
        &00100000 \longrightarrow R_X \left( -\Theta \right), \\
        &00001000 \longrightarrow R_Y \left( -\Theta \right), \\
        &00000010 \longrightarrow R_X \left( \Theta \right).
    \end{aligned}
\label{eq:pulse_position_rotations}
\end{equation}

In the general case, if a qubit period is divided into $n$ equal parts by choosing $\omega_g = n \omega_{01}$, a pulse applied at the $k$-th position will result in the qubit operation $\hat{U}_{nk}$, defined by formula:
\begin{equation}
\hat{U}_{nk} \left( \Theta \right) = R_Y \left( \Theta \right)  R_Z \left( 2 \pi  \frac{k-1}{n} \right).
\label{eq:pulse_slot_operation}
\end{equation}

For the regular pattern $(10000000)^M$, where $M$ is the number of regular pulses, the nonzero SFQ pulse occurs at the same phase of each qubit period. In the ideal two-level approximation, each repetition therefore produces the same rotation $R_Y(\Theta)$, so that
\begin{equation}
    \left[R_Y(\Theta)\right]^M  =  R_Y({M}\Theta).
\label{eq:regular_rotation_accumulation}
\end{equation}

For $\Theta=0.030$, choosing $M=52$ gives $M \Theta = 1.56~\mathrm{rad} \simeq \pi/2$. Thus, the regular sequence provides the natural starting point for constructing the target $Y_{\pi/2}$ gate. The complete multilevel evolution, including leakage and phase errors, is evaluated numerically rather than inferred from this two-level estimate.

\begin{figure*}[t]
    \centering
    \includegraphics[width=\textwidth]{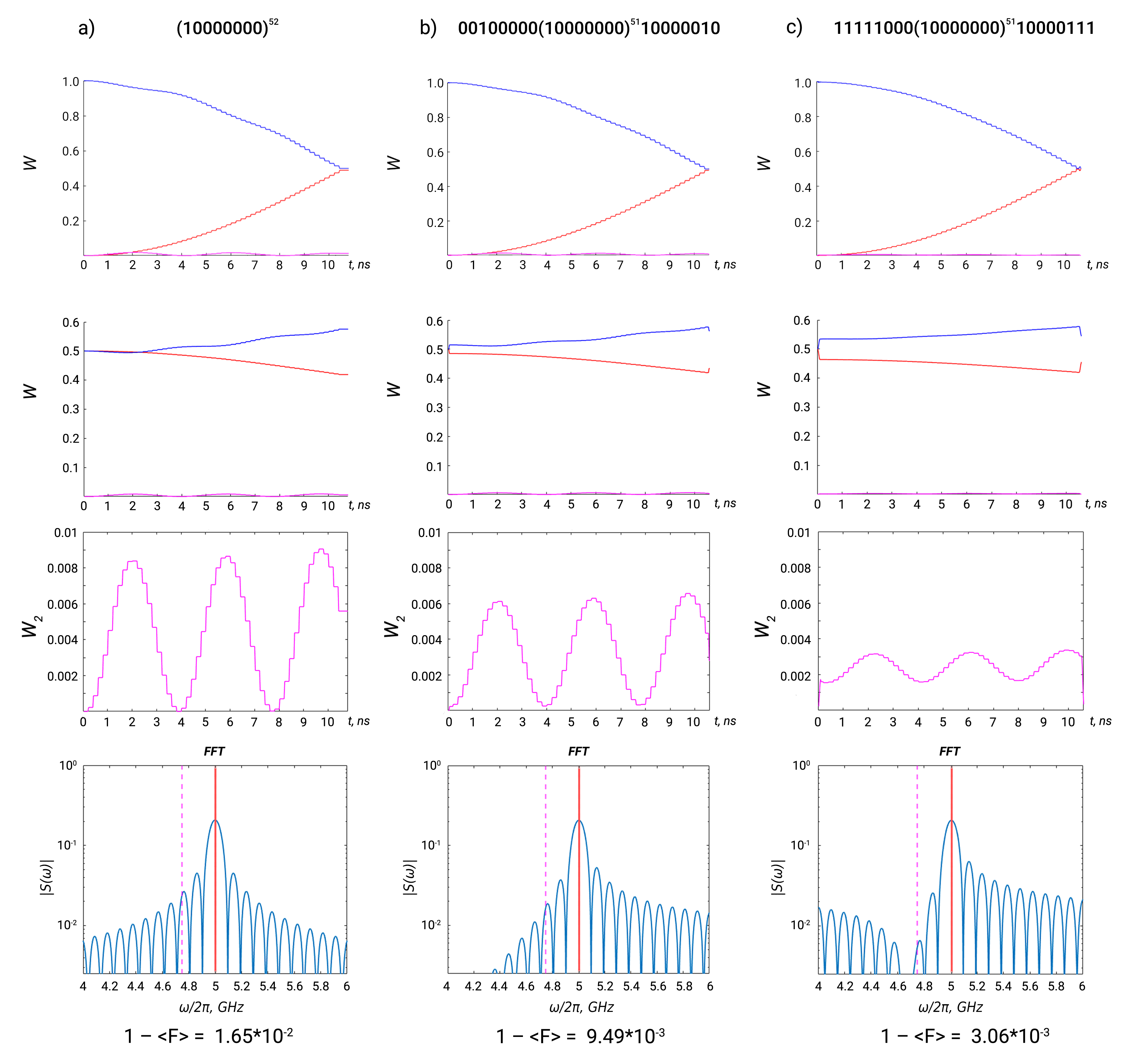}
    \caption{
    Comparison of three SFQ control schemes for the target
    $Y_{\pi/2}$ gate:
    (a) regular sequence without ramps,
    (b) combined sequence with symmetric ramps and
    (c) combined sequence with independently optimized ramps.
    First row: evolution from the initial state $\ket{z+}=\ket{0}$, illustrating the intended $\pi/2$ rotation through the evolution toward equal $\ket{0}$ and $\ket{1}$  populations.
    Second row: evolution from the initial state $\ket{y+}$. Since $\ket{y+}$ is an eigenstate of the $Y$ rotation axis, an ideal $Y_{\pi/2}$ gate leaves its computational-basis populations equal to $1/2$; deviations therefore highlight the additional effects of the control sequence and, in panels (b) and (c), the
    short ramp corrections.
    Third row: population $W_2$ of the first leakage level for the $\ket{y+}$ initial state.
    Bottom row: spectra of the corresponding SFQ control signals. The solid vertical line indicates the qubit transition frequency $\omega_{01}$ and the dashed vertical line the leakage transition frequency $\omega_{12}$. The infidelity shown below each column is the leakage-inclusive average gate infidelity calculated over all six cardinal input states according to Eq.~\eqref{eq:fidelity}, rather than from either of the individual trajectories displayed here.
    The rapid changes near the boundaries of panels (b) and (c) correspond to the short on- and off-ramp sequences generated on the $40~\mathrm{GHz}$ timing grid.
    }
    \label{fig:2}
\end{figure*}

Figure~\ref{fig:2} compares the three control schemes using two representative initial states. The evolution from $\ket{z+}=\ket{0}$ is used to visualize the intended $Y_{\pi/2}$ rotation: ideally, $Y_{\pi/2}\ket{0}$ produces equal populations of $\ket{0}$ and $\ket{1}$. The evolution from $\ket{y+}$ provides a complementary view of the control errors. Since $\ket{y+}$ is an eigenstate of the $Y$ operator, an ideal $Y_{\pi/2}$ operation changes it only by a global phase and therefore leaves its computational-basis populations equal to $1/2$. Deviations from this behavior make the effects of the ramp corrections particularly visible.

The leakage dynamics are shown separately for the $\ket{y+}$ initial state, while the bottom panels show the spectra of the corresponding SFQ control signals. The infidelity reported below each panel characterizes the complete $Y_{\pi/2}$ gate and is calculated from all six cardinal input states according to Eq.~\eqref{eq:fidelity}; it is not calculated from either of the individual trajectories shown in the figure.

\begin{description}
\item [Sequence with no optimization of the ramps] \hfill \\

We first consider the regular sequence $(10000000)^{52}$. As discussed above, its nominal accumulated rotation is $52\Theta=1.56~\mathrm{rad}\simeq\pi/2$. The $\ket{z+}$ trajectory in Fig.~\ref{fig:2}(a) shows the corresponding evolution toward approximately equal populations of $\ket{0}$ and $\ket{1}$, as expected for a $Y_{\pi/2}$ operation. For the $\ket{y+}$ input, deviations from the ideal $W_0 = W_1 = 1/2$ behavior arise from leakage and coherent errors. The leakage-inclusive average gate infidelity is $1-\braket{F}=0.0165$.

\item [Sequence with symmetric optimization of the ramps] \hfill \\
We use the idea from \citet{shillito} which involves bracketing a regular sequence with symmetrical ramps. These ramps should be symmetric with respect to Y-rotations and antisymmetric with respect to X-rotations. As an example, consider the following sequence:
\begin{equation}
        00100000 \left( 10000000 \right)^{{M}-1} 10000010.
\label{eq:symmetric_sequence_example}
\end{equation}

In terms of qubit operations it is equal to
\begin{equation}
    R_X \left( \Theta \right) R_Y \left( \Theta \right)^{M}  R_X \left( -\Theta \right).
\label{eq:symmetric_sequence_operator}
\end{equation}

The effect of the short boundary corrections can also be seen in the $\ket{y+}$ trajectory in Fig.~\ref{fig:2}(b), particularly near the beginning and end of the sequence. The corresponding leakage is reduced compared with the regular sequence, and the average gate infidelity decreases to $1-\braket{F}=0.00949$.

\item [Sequence with unrestricted optimization] \hfill \\
As the size of the control sequence is small, the optimal sequence can be found by a simple exhaustive sweep. There are only 65536 possible subsequences for the ramps (256 for the on-ramps and 256 for the off-ramps). The sweep shows that the highest fidelity is achieved with the sequence 11111000 for the on-ramp and 10000111 for the off-ramp. In terms of qubit operations, this corresponds to the following sequence:
\begin{eqnarray}
    R_Y \left( -\Theta \right) R_Z \left( \pi / 4 \right) R_X \left( -2 \Theta \right) R_Z \left( \pi / 4 \right)  R_Y \left( \Theta \right)^{M+2}  \label{eq:unrestricted_sequence_operator}\\ R_Z \left( \pi / 4 \right) R_X \left( 2 \Theta \right) R_Z \left( \pi / 4 \right) \nonumber.
\end{eqnarray}

The independently optimized ramps provide a further reduction of both leakage and total gate error, as shown in Fig.~\ref{fig:2}(c). The resulting average gate infidelity is $1-\braket{F}=0.00306$.

We compare the results of the three types of sequence implementations for $Y_{\pi/2}$ gate in Fig.~\ref{fig:2}. The regular (central) part of the sequence was generated at the qubit frequency, $\omega^{(reg)}_{g} = \omega_{01}$, while the ramps were generated at a frequency much higher than the main qubit frequency, $\omega^{(ramp)}_g/2\pi = 8\,\omega_{01}$. We examine not only the qubit-level evolution to verify that the desired rotation was correctly enacted but also the spectrum of the SFQ control pulse. The spectra provide a qualitative interpretation of the leakage reduction: reduced spectral weight near $\omega_{12}$ is associated with suppression of the $\ket{1} \rightarrow \ket{2}$ transition whereas the genetic algorithm itself optimizes the complete gate fidelity.”
\end{description}

\subsection{Sequence search methodology and results}\label{subsec:sequence_search}

Since the optimal solutions cannot be inferred solely from symmetry principles, optimal search algorithms that are free from such constraints can be employed. In our previous work, we have used gradient descent and genetic algorithms, with the latter yielding the most promising results \citep{genetic_algo}. Finding the optimal combined sequence is equivalent to identifying the best ramps for a given set of parameters, including the length of the regular part of the sequence.

Here, we demonstrate the possibility of obtaining optimal combined sequences using a genetic algorithm. This approach is particularly suitable for this task, as SFQ qubit control sequences are inherently discrete. During the search process, the algorithm employs two forms of permutations: \textit{crossover}, in which a crossover point is selected and the corresponding trailing segments of two parent sequences are exchanged,, and \textit{mutations}, which randomly alter individual symbols within the sequence. Crossover allows extensive exploration of the control landscape by radically changing the sequence, while mutations make small incremental steps, exploring only sequences neighboring the current one. Additional details on the working principle of the genetic algorithm can be found in \citet{genetic_algo}. We used the following hyperparameters: mutation probability $P_m = 0.5$, crossover probability $P_c = 0.8$.  One-point crossover was applied with probability $P_c$. Mutation was applied to each offspring with probability $P_m$; conditional on mutation, each sequence element was modified with probability $1/N$. The two highest-fidelity individuals were retained between generations. The algorithm was run for $MaxIter = 500$. The population size was not fixed independently but was chosen as twice the sequence length. For each parameter set, 96 independent genetic-algorithm runs were performed using different random initial populations. The sequence with the highest gate fidelity among the 96 runs was retained as the best-fidelity solution. The computational cost estimation is discussed in \citet{genetic_algo} as the same machine was used in this article.

\begin{figure}[htbp]
    \centering
    \includegraphics[width=1\linewidth]{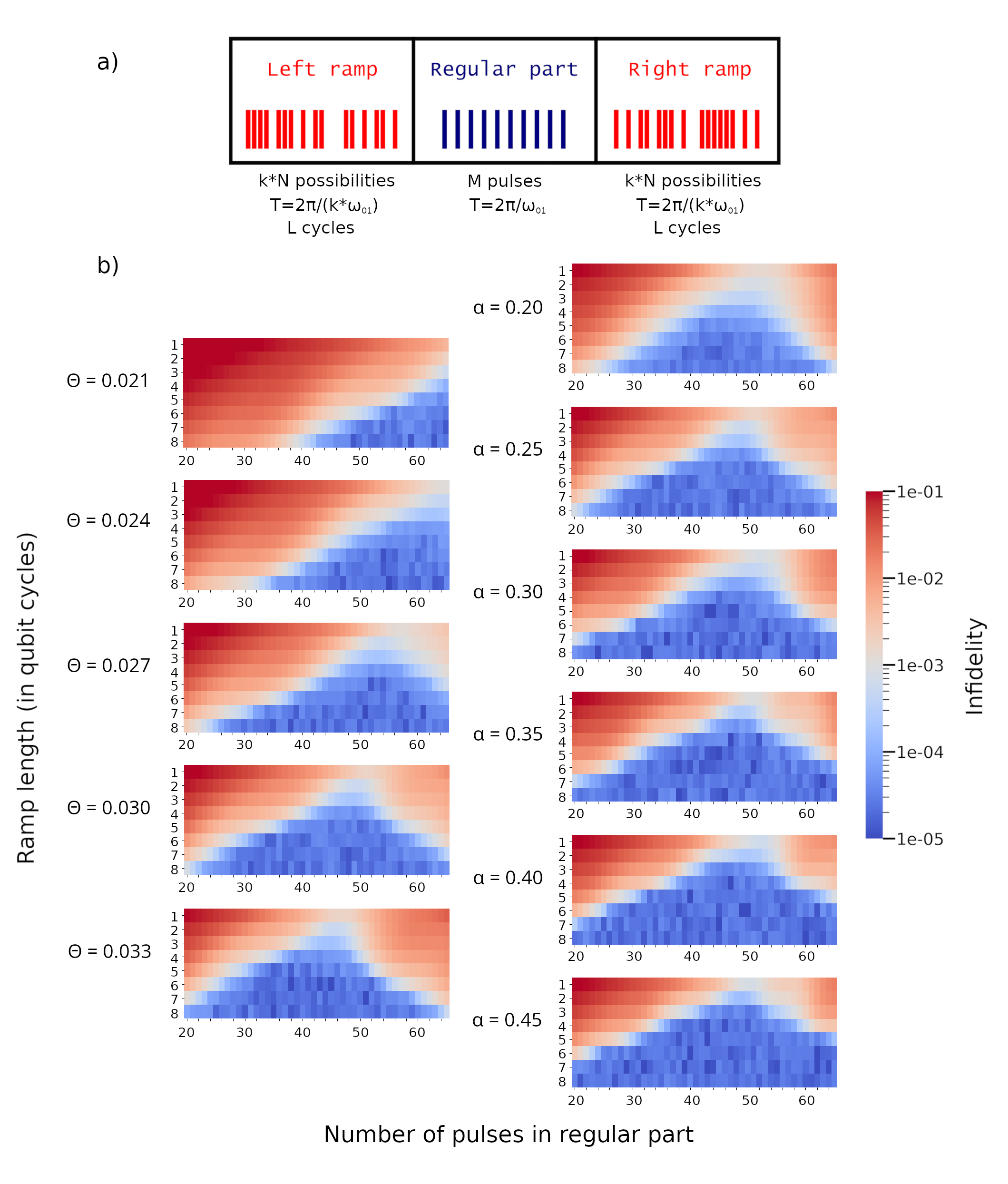}
    \caption{Search results using genetic algorithm for combined  sequences with different qubit parameters: anharmonicities $\alpha$ and single pulse qubit rotation angles $\Theta$. The main qubit frequency was chosen as $\omega_{01}/2\pi = 5$~GHz, the generator frequency $\omega_g/2\pi = 40$~GHz, and the pulse width as $\tau = 4$ ps. (a) Combined sequence structure scheme. (b) Search results. Each point in the plot indicates the infidelity of the best sequence found using the genetic algorithm. On the y-axis, the number of pulses is expressed in qubit free precession cycles, with each cycle containing $k = \omega_g / \omega_{01}=8$ pulses.
    }
    \label{fig:3}
\end{figure}

We have performed the search algorithm for the system with different parameters: qubit anharmonicities $\alpha$ ranging from 0.2 to 0.45~GHz and rotation angles $\Theta$ ranging from 0.021 to 0.033. The results of the search can be seen on Fig.~\ref{fig:3}. 

For every parameter set considered, the genetic algorithm identifies a complete SFQ sequence implementing the target gate $Y_{\pi/2}$ with coherent-control infidelity $1-\braket{F}<10^{-4}$. Here and throughout this section, the reported infidelity includes leakage from the computational subspace but excludes relaxation, dephasing, and other implementation-dependent incoherent errors. The genetic algorithm is sufficiently powerful to identify high-fidelity sequences even for small anharmonicities. The results for the different parameter sets are summarized in Table~\ref{tab:performance}. In all optimization runs presented in the table, the qubit frequency is fixed at $\omega_{01}/2\pi = 5~\mathrm{GHz}$.

For each pair of $\alpha$ and $\Theta$, several different criteria are used to characterize the obtained population of control sequences. The values of $1-\braket{F}$ and leakage $L$ correspond to the single sequence with the highest fidelity found in the search. By contrast, the reported memory requirement $N$ is the minimum number of ramp bits among sequences satisfying $1-\braket{F} < 10^{-4}$, while $t_{\mathrm{gate}}$ is the shortest gate time found under the same fidelity constraint. Consequently, the best-fidelity sequence, the minimum-memory sequence, and the shortest-duration sequence are generally different and the corresponding values in Table~\ref{tab:performance} should not be interpreted as simultaneous characteristics of a single control sequence.

For the nominal parameter set $\alpha/2\pi=0.25~\mathrm{GHz}$ and $\Theta=0.030$ (parameter set \#2 in Table~\ref{tab:performance}), the minimum-memory sequence satisfying $1-\braket{F}<10^{-4}$ has $ M=48,\quad N=64$, where $M$ is the number of pulses in the regular part and $N$ is the number of stored ramp bits. This sequence is used below as the baseline configuration for the controller-complexity estimate in Sec.~\ref{subsec:logical_complexity}.

Increasing the single-pulse rotation angle enables reductions in memory requirement and gate duration while maintaining coherent-control infidelity below $10^{-4}$.  Symmetry analysis of the  sequences once again shows that the symmetry of $X$- and $Y$-operations in the ramps does not play a significant role. The results of studying the influence of symmetry on the performance of the genetic algorithm are provided in Appendix~\ref{app:symmetry}.

\begin{table}[htp!]
\caption{
Performance of combined sequences found using the genetic algorithm for  different anharmonicities $\alpha$ and single-pulse rotation angles $\Theta$. The qubit frequency is fixed at $\omega_{01}/2\pi = 5~\mathrm{GHz}$ for all parameter sets. Each row summarizes the results obtained for one parameter set from
96 independent genetic-algorithm runs. The reported performance metrics do not necessarily correspond to the same control sequence. The infidelity $1-\braket{F}$ and leakage $L$ are reported for the sequence with the highest fidelity found in the corresponding search. The memory requirement $N$ denotes the smallest number of ramp bits among sequences with $1-\braket{F} < 10^{-4}$, while $t_{\mathrm{gate}}$ is the shortest gate time among sequences satisfying the same fidelity criterion. Thus, the best-fidelity, minimum-memory, and shortest-duration results may correspond to different sequences.
}
\label{tab:performance}

\centering
\begin{tabular}{|c|c|c|c|c|c|c|}
    \hline
    $\#$ & $\alpha/2\pi$,~GHz & $\Theta$  &$L, 10^{-5}$ & $1-\braket{F}, 10^{-5}$ & min $N$ & min $t_{gate}$,~ns\\
    \hline
    1 & 0.20 & 0.030 & 0.29 & 1.46 & 80 & 9\\
    \hline
    2 & 0.25 & 0.030 & 0.63 & 1.25 & 64 & 7.8\\
    \hline
    3 & 0.30 & 0.030 & 0.07 & 0.39 & 48 & 7.2\\
    \hline
    4 & 0.35 & 0.030 & 0.31 & 0.53 & 48 & 7.2\\
    \hline
    5 & 0.40 & 0.030 & 0.18 & 0.39 & 48 & 7.2\\
    \hline
    6 & 0.45 & 0.030 & 0.19 & 0.37 & 48 & 6.8\\
    \hline
    7 & 0.25 & 0.021 & 0.67 & 1.28 & 80 & 11.8\\
    \hline
    8 & 0.25 & 0.024 & 0.47 & 0.66 & 64 & 10.2\\
    \hline
    9 & 0.25 & 0.027 & 0.32 & 0.66 & 64 & 8.4\\
    \hline
    10 & 0.25 & 0.033 & 0.66 & 0.94 & 64 & 7.2\\
    \hline
\end{tabular}
\end{table}

It should be emphasized that Fig.~\ref{fig:3} demonstrates the ability to identify a high-fidelity sequence separately for each parameter set and does not establish robustness of a fixed sequence. The robustness of one representative optimized sequence against parameter variations (qubit frequency $\omega_{01}$, anharmonicity $\alpha$ and single-pulse rotation angle $\Theta$) and timing jitter is analyzed in Appendix~\ref{app:robustness}.

\subsection{Memory optimization}\label{subsec:memory_optimization}

\begin{figure}[htbp]
    \centering
    \includegraphics[width=1\linewidth]{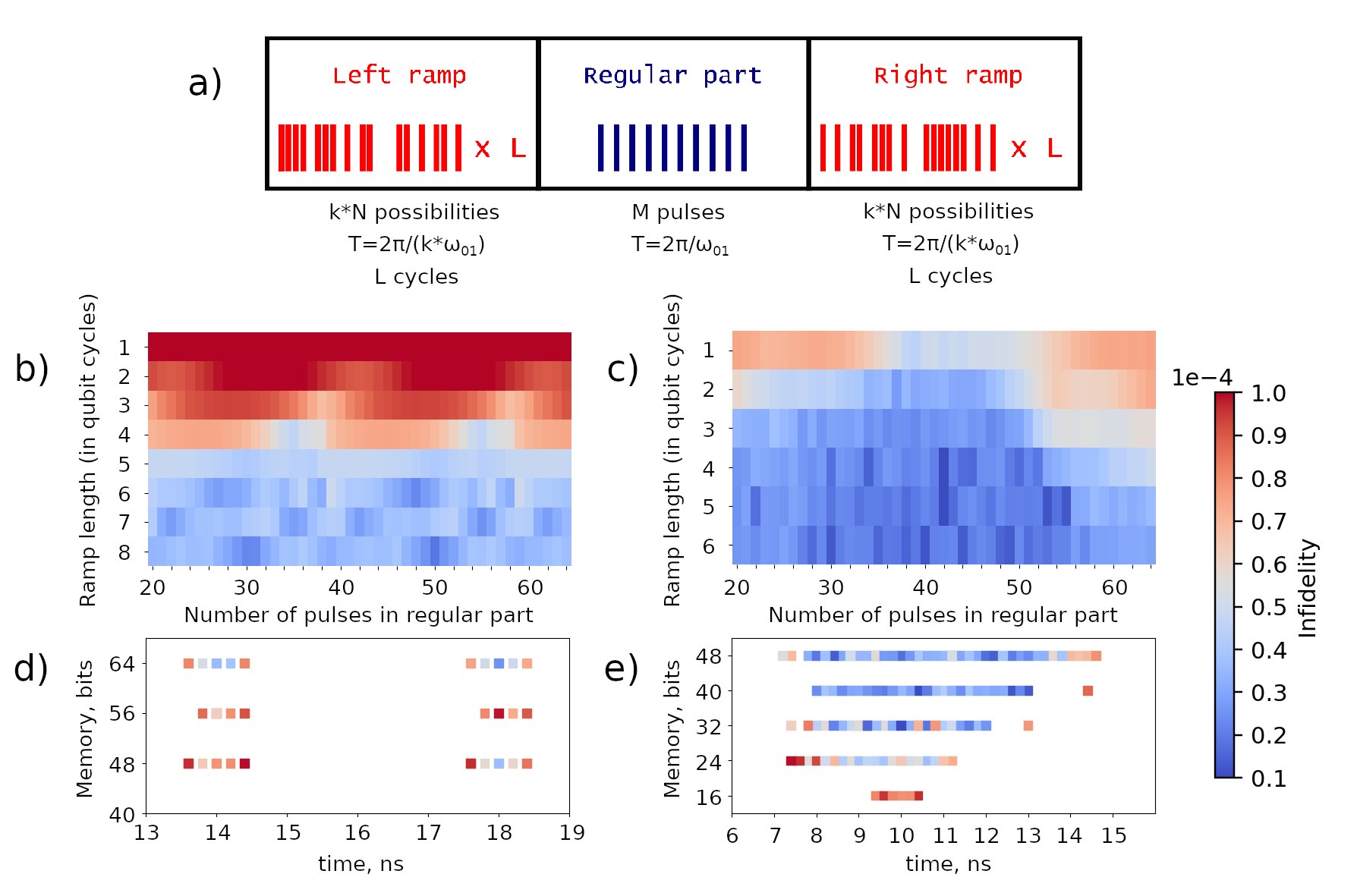}
    \caption{Search results using a genetic algorithm for effective qubit control SFQ sequences in the form of repeating subsequences. Search parameters: main qubit frequency $\omega_{01}/2\pi = 5$~GHz, anharmonicity $\alpha/2\pi = 0.25$~GHz, single-pulse rotation angle $\Theta = 0.030$, pulse width $\tau = 4$ ps, and generator frequency $\omega_{g}/2\pi = 40$~GHz. (a) Combined sequence structure using subsequences. In plots (b) and (d), the results are shown for the case, where the left and right ramps were searched as identical, while in (c) and (e), the ramps were optimized independently. (b)--(c): Visualization of the search results, indicating the lowest infidelity value found for a fixed number of pulses in the regular part of the sequence and in the arbitrary part, expressed in qubit free precession cycles, each cycle containing $k=\omega_g / \omega_{01}=8$ pulses. The number of repetitions $L$ is not shown on the plot. (d)--(e) Visualization of the memory/time efficiency of the sequences.
    }
    \label{fig:4}
\end{figure}

The results suggest that a considerable amount of memory is still required to store the control sequences, necessitating the use of 64-bit or 128-bit memory cells for sequence storage. One approach to reduce the required memory is described in \citet{sfq_leakage}: the control sequence can be represented as a series of small repeating subsequences. We adapted the genetic algorithm to search for combined sequences in which each ramp is repeated $L$ times. Two cases were considered: (1) the left and right ramps are identical (allowing them to be stored simultaneously), and (2) the left and right ramps are optimized independently.

As shown in Fig.~\ref{fig:4}, both implementations of the combined sequence scheme substantially reduce the required memory. In the equal ramp configuration, the minimum-memory solution requires 48 bits of memory. Allowing the left and right ramps to be optimized independently further reduces the memory requirement to 16 bits.

For the same nominal qubit and pulse parameters, the repeating-subsequence search identifies a minimum-memory solution with
\begin{equation}
    M=37,\qquad N=16,\qquad L=5,
\label{eq:repeating_memory_solution}
\end{equation}

where $L$ is the number of repetitions of the stored ramp subsequence. Thus, the stored ramp length is reduced from 64 to 16 bits compared with the baseline sequence from Table~\ref{tab:performance}.

The sequence achieving the minimum memory requirement is not the same as the sequence with the shortest gate duration. The shortest gate time obtained in the independently optimized case is 9.4~ns, illustrating the trade-off between memory usage and gate duration.

\section{Qubit controller scheme}\label{sec:controller}

This section presents a conceptual SFQ controller architecture capable of realizing the regular and arbitrary components of the combined control sequence. The purpose is to identify the required functional modules and estimate their logical complexity using standard SFQ building blocks. 

We can outline the key characteristics that a capable qubit controller must possess to implement the combined sequence control scheme. First, it must support both forms of control: regular and arbitrary. Second, the controller should follow a modular design philosophy, with each module performing a well-defined function. Third, the controller must be scalable with respect to both the memory capacity of its individual components and the total number of modules. In this work, we introduce two primary controller modules: the Regular Sequence Generator (RSG) and the Arbitrary Sequence Generator (ASG), which realize the corresponding types of SFQ qubit control. All controller subsystems can be assembled from standard SFQ building blocks: T flip-flops (TFF), D flip-flops (DFF), and Non-Destructive Readout (NDRO) cells.

\subsection{Regular Sequence Generator}\label{subsec:rsg}
\begin{figure}[htbp]
    \centering
    \includegraphics[width=0.8\linewidth]{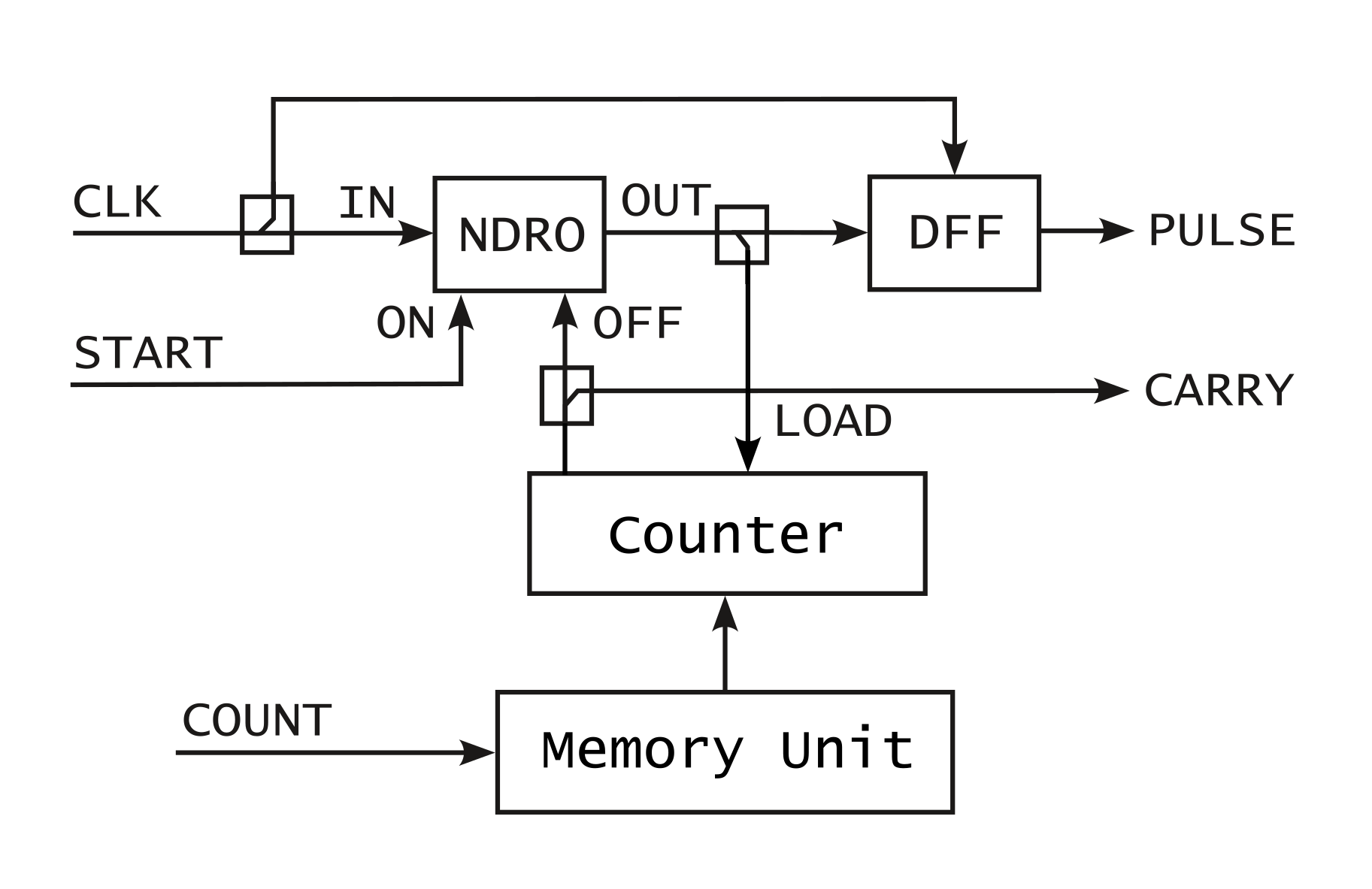}
    \caption{Schematic of the Regular Sequence Generator (RSG).}
    \label{fig:5}
 \end{figure}

The Regular Sequence Generator (RSG) receives an external periodic SFQ clock signal and produces a finite train containing a prescribed number $M$ of SFQ pulses. The generation process begins when a signal is applied to the START input of the NDRO cell, which functions as a controllable switch. To achieve precise control over the number of generated SFQ pulses, an SFQ counter is used following an approach inspired by \citet{controller_design6}. The counter is constructed as a series of T flip-flops (TFFs) with parallel input and serial output. Once the number of generated pulses reaches the target value $M$, a signal is sent to the OFF input of the NDRO switch, halting further pulse propagation. The counter can be preconfigured by loading the required count through the COUNT channel. For example, a counter consisting of 4 TFFs with states “1” recorded in the third and fourth counters will produce $2^4 - 2 - 1 = 13$ pulses. SAt the completion of the counting process, the counter also generates a CARRY signal that indicates the end of the regular sequence and can be used to initiate the subsequent controller module. Additionally, a D flip-flop placed at the end of the RSG module functions as a clock synchronizer for the output sequence. Fig.~\ref{fig:5} illustrates the schematic of a regular sequence generator. Memory scalability can be achieved by increasing both the number of flip-flops and the size of the corresponding memory units.

\subsection{Arbitrary Sequence Generator}\label{subsec:asg}
The Arbitrary Sequence Generator can be constructed using DFF-based shift registers and TFF-based counters. The DFF-based shift register receives the sequence from the memory unit, with data transfer occurring asynchronously before the sequence is executed. Simultaneously, information about the number of sequences is sent to the counter in the same manner. Similar to the RSG, the sequence start is controlled by an NDRO cell functioning as a switch. The START signal opens the switch, allowing the clock signal (CLK) to reach both the counter and the shift register. At the end of the sequence, the counter triggers the OFF input of the NDRO switch, halting the sequence generator and sending a CARRY signal to the next module.

\begin{figure}[htbp]
    \centering
    \includegraphics[width=1\linewidth]{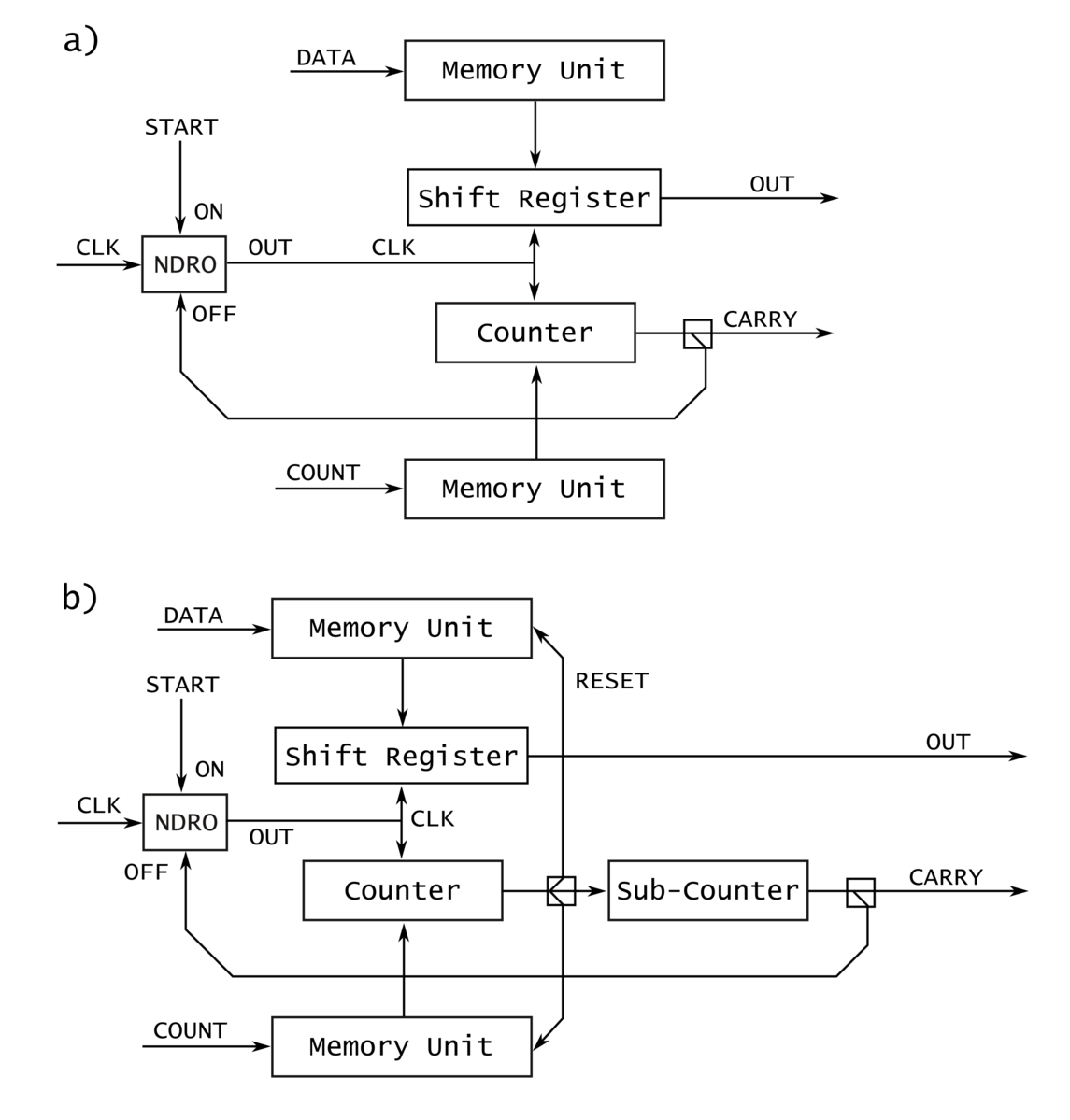}
    \caption{Schematic of the Arbitrary Sequence Generator (ASG). (a) Implementation of the control protocol without repetitions. (b) Implementation of the control protocol with repetitions.}
    \label{fig:6}
 \end{figure}

In the combined sequence approach, which uses two arbitrary sequences (such as ramps), both the counter and the shift register must be reset during execution of the regular sequence. Fig.~\ref{fig:6}(a) depicts the layout of the arbitrary sequence generator. Previously, we examined scenarios in which arbitrary sequences can be constructed as repeating subsequences. To implement this, the existing controller architecture must be modified by replacing the single counter with a serial pair of counters. Once the first counter completes its cycle, indicating that the current subsequence has finished, its output pulse increments the second counter (referred to as the sub-counter), which tracks how many times the subsequence is to be repeated. Simultaneously, with every tick of the second counter, the data defining the subsequence is reloaded from the memory block into the first counter and the shift register. Alternatively, a ring counter may be used in place of a conventional shift register. In this configuration, the entire module’s operation, including generation of the CARRY signal, terminates when the second counter has completed its count. This modified scheme is shown in Fig.~\ref{fig:6}(b).

\subsection{Combined sequence modular controller}\label{subsec:modular_controller}

\begin{figure}[htbp]
    \centering
    \includegraphics[width=1\linewidth]{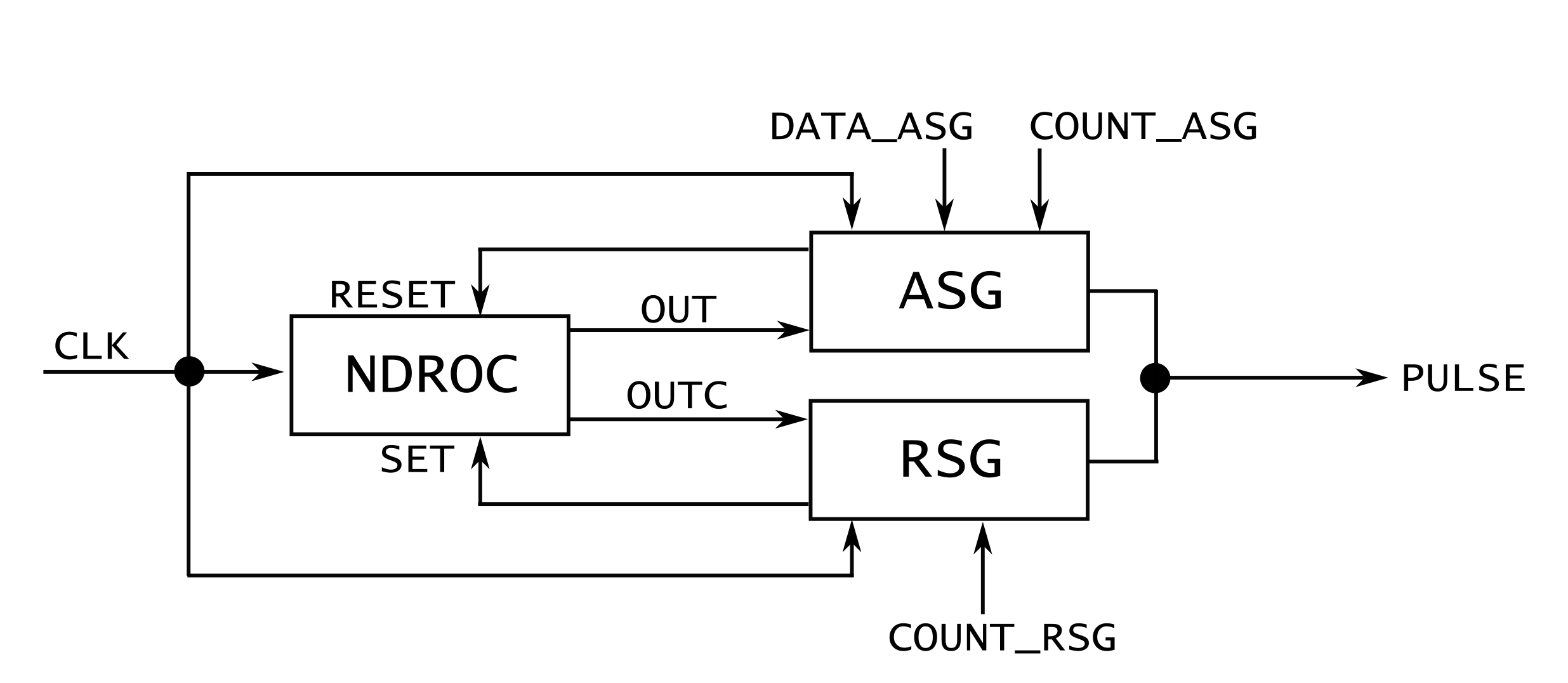}
    \caption{Full controller schematic showing the connection of RSG and ASG modules in order to realize the DRAG-inspired approach of the qubit control.}
    \label{fig:7}
 \end{figure}

Fig.~\ref{fig:7} illustrates how the RSG and ASG modules are integrated into the overall qubit SFQ controller. The modules are interconnected such that the completion of one module’s sequence triggers the next, following a protocol in which a regular sequence is framed by two arbitrarily composed ramps. Transitions between modules are controlled by an NDROC cell operating as a switch \citep{ndroc}. By toggling its state via the SET/RESET inputs, either OUT or OUTC is selected as the active output in alternation. Each module generates a CARRY signal that drives the NDROC cell, enabling the transition to the other module. Both modules can access a shared memory unit—such as an NDRO-based register, a cryogenic RAM capable of addressing SFQ cells, or an SFQ-based RAM unit. Because the modules never run concurrently, the stored data can be overwritten while sequences are being executed, reducing memory requirements. Alternatively, the NDRO switches within the modules could be removed entirely by routing START signals directly to the memory unit; however, this would require additional clock synchronization with the controller.

\subsection{Logical complexity estimation}\label{subsec:logical_complexity}

The logical complexity of the proposed controller can be estimated from the number of Josephson junctions (JJs) required by its constituent SFQ cells. According to the circuit implementations used here, a D flip-flop (DFF) 5  JJs~\citep{sfq_bakolo}, a T flip-flop (TFF) 8 JJs~\citep{sfq_bakolo}, an NDRO cell 7 JJs~\citep{sfq_polonsky} and an NDROC cell 12 JJs~\citep{ndroc}. The following estimate is intended only to characterize the logical complexity of the controller architecture. It does not constitute a physical area or power estimate.

For a regular sequence containing $M$ pulses, the RSG requires
\begin{equation}
    n_M = \left\lfloor\log_2 M\right\rfloor + 1
\label{eq:rsg_counter_bits}
\end{equation}

TFFs for the pulse counter, together with one NDRO cell and one DFF.
The corresponding junction count is therefore
\begin{equation}
    JJ_{\mathrm{RSG}} = 8 n_M + 7 + 5.
\label{eq:rsg_jj_count}
\end{equation}

For an arbitrary ramp containing $N$ stored bits, the ASG requires
$N$ DFFs in the shift register,
\begin{equation}
    n_N = \left\lfloor \log_2 N \right\rfloor + 1
\label{eq:asg_counter_bits}
\end{equation}

TFFs in the counter, and one NDRO cell. Its logical junction count is
\begin{equation}
    JJ_{\mathrm{ASG}} = 5N + 8n_N + 7.
\label{eq:asg_jj_count}
\end{equation}

One additional 12-JJ NDROC element is used to switch between the RSG and ASG modules. Since the left and right ramps are not executed simultaneously, the same ASG hardware can be reused for both ramps; only one ramp needs to be stored at a time. The total junction count for the baseline combined-sequence controller is therefore
\begin{equation}
\begin{split}
    JJ_{\mathrm{base}} &= JJ_{\mathrm{RSG}} + JJ_{\mathrm{ASG}} + 12 \\
    &= 8\left(\left\lfloor\log_2 M \right\rfloor+1\right) + 8\left(\left\lfloor\log_2 N \right\rfloor+1\right) + 5N + 31.
\end{split}
\label{eq:jj_base}
\end{equation}

To provide a representative estimate, we use parameter set \#2 of Table~\ref{tab:performance}, corresponding to $\alpha/2\pi=0.25~\mathrm{GHz}$ and $\Theta=0.030$. For this parameter set, the minimum-memory sequence satisfying $1-\braket{F}<10^{-4}$ has $M=48$ regular pulses and $N=64$ stored ramp bits. This is the same sequence identified in Sec.~\ref{subsec:sequence_search} as the minimum-memory solution for this parameter set.

For $M=48$ and $N=64$,
\begin{equation}
    n_M = 6, \qquad n_N = 7,
\label{eq:baseline_counter_bits}
\end{equation}

and Eq.~\eqref{eq:jj_base} gives
\begin{equation}
    JJ_{\mathrm{base}} = 60 + 383 + 12 = 455.
\label{eq:baseline_jj_count}
\end{equation}

We next consider the repeating-subsequence optimization introduced in Sec.~\ref{subsec:memory_optimization} for the same qubit and pulse parameters. A minimum-memory solution obtained in this search is characterized by
\begin{equation}
    M=37, \qquad N=16, \qquad  R=5,
\label{eq:repeating_controller_solution}
\end{equation}

where $N$ is now the length of the stored ramp subsequence and $R$ is the number of times this subsequence is repeated. Compared with the baseline controller, this implementation requires one additional TFF counter in the ASG to store the repetition number $R$. The corresponding contribution is
\begin{equation}
    JJ_{\mathrm{sub}} = 8 \left(\left\lfloor \log_2 R \right\rfloor +  1 \right).
\label{eq:repetition_counter_jj}
\end{equation}

The total junction count for the repeating-subsequence controller is therefore
\begin{equation}
    JJ_{\mathrm{rep}} = JJ_{\mathrm{base}}(M,N) + JJ_{\mathrm{sub}},
\label{eq:repeating_jj_total}
\end{equation}

where $JJ_{\mathrm{base}}(M,N)$ is evaluated using the sequence parameters of the repeating-subsequence solution. For $M=37$, $N=16$, and $R=5$,
\begin{equation}
    n_M=6, \qquad n_N=5, \qquad n_R=3,
\label{eq:repeating_counter_bits}
\end{equation}

which gives
\begin{equation}
\begin{split}
    JJ_{\mathrm{rep}} &= 60 + 127 + 12 + 24 = 223.
\end{split}
\label{eq:repeating_jj_count}
\end{equation}

Thus, for the same nominal qubit and pulse parameters, the repeating-subsequence optimization reduces the stored ramp length from 64 to 16 bits and the estimated logical junction count from 455 to 223, corresponding to a reduction of approximately 51\%. The reduction in the total junction count is smaller than the fourfold reduction in stored ramp length because the RSG, counters, and switching elements introduce a fixed logical overhead that is largely independent of the ramp-memory size. These values should be interpreted as estimates of the relative logical complexity of the two controller architectures rather than as complete circuit-level resource estimates. 

\subsection{Clocking, synchronization and scope of the controller architecture}\label{subsec:clocking}

The controller schemes in Figs.~\ref{fig:5}--\ref{fig:7} represent an  architecture concept rather than a complete circuit-level implementation. Although the proposed modules are composed of standard SFQ logic elements, a physical implementation additionally requires a process-specific clock distribution network, Josephson transmission lines, splitter trees, bias circuitry and layout-dependent interconnects. No circuit-level timing-margin or parameter-margin analysis of the complete controller is performed in the present work.

A particularly important implementation issue is synchronization of the regular and arbitrary parts of the combined sequence. For the representative parameters considered here, the ramp sequence uses a clock frequency equal to $\omega_{g}/2\pi = 40~\mathrm{GHz}$, corresponding to a $25~\mathrm{ps}$ timing interval, whereas the regular part is synchronized to the nominal qubit frequency $\omega_{01}/2\pi = 5~\mathrm{GHz}$. Since $\omega_{g}/\omega_{01} = 8$, the two timing references can in principle be derived from a common master reference, such that regular sequence pulses occur on every 8th edge of the high-frequency timing grid. 

Transitions between the ASG and RSG modules introduce an additional timing constraint. In the conceptual architecture, the completion of one module generates a CARRY signal that enables the following module. In a physical implementation, the propagation delay through the counter, switching, clock distribution and interconnect networks must be accounted for. A fixed, deterministic delay between controller modules can be included directly in the sequence timing. In contrast, pulse-to-pulse jitter and variations in the relative clock skew are not deterministic and therefore require a separate timing-margin analysis. One possible implementation is to register the module-transition signal with the common clock, so that the next sequence generator is enabled only on a defined clock edge. A conceptually similar approach is illustrated in Fig.~6 of \citet{quantum-classical}, where an additional synchronization circuit is used to coordinate the transition between modules.

The minimum-memory result discussed in Sec.~\ref{subsec:memory_optimization} requires only a single 16-bit ramp register, even though the independently optimized left and right ramps are different. In this implementation, the register is initially loaded with the subsequence defining the left ramp. After the left ramp has been generated and control is transferred to the RSG, the same register can be overwritten with the subsequence defining the right ramp while the regular pulse train is being executed. The ASG is then reused to generate the right ramp after completion of the regular sequence. Thus, the 16-bit memory requirement refers to the maximum ramp information that must be stored simultaneously, rather than to the total number of bits describing both ramps. This implementation requires the memory reload to be completed before the end of the regular sequence; the corresponding write latency therefore constitutes an additional timing constraint for a physical controller implementation.

The parameter robustness analysis of Appendix~\ref{app:robustness} provides a control-level estimate of the timing precision required by the optimized sequence. For the representative sequence considered there, the mean coherent-control infidelity remains below $10^{-4}$ for rms independent pulse-placement jitter up to approximately $1.32~\mathrm{ps}$, while at least 95\% of Monte Carlo realizations satisfy the same criterion for $\sigma_t\leq0.7~\mathrm{ps}$. These values should be interpreted as qubit-level timing tolerances rather than demonstrated circuit timing margins.

Synchronization to the actual qubit frequency provides a second constraint. The same robustness analysis shows that, with the $40~\mathrm{GHz}$ clock kept fixed, the representative sequence satisfies $1-\braket{F}<10^{-4}$ only within a sub-MHz interval around the nominal $5~\mathrm{GHz}$ qubit frequency. Experimental operation would therefore require calibration of the qubit frequency and, when necessary, retuning of the timing reference or reoptimization of the stored sequence.

The numerical qubit evolution is calculated using the complete time-dependent driven Hamiltonian, so deterministic phase accumulation associated with the applied SFQ drive is included in the optimized evolution. The proposed controller does not, however, implement real-time tracking of drive-induced frequency shifts. Device-specific calibration of the effective driven qubit response and subsequent sequence reoptimization may therefore be required.

A complete timing closure of the proposed architecture would require an architectural timing model including cell latencies, setup constraints, clock skew and interconnect delays, followed by process-specific circuit and layout simulation. Such an analysis is left for future work.

\section{Conclusions}\label{sec:conclusions}

This work presents a method for constructing memory-efficient SFQ control sequences for superconducting transmon qubits. Using a genetic-algorithm-based search, we demonstrate representative $Y_{\pi/2}$ sequences with coherent-control infidelities below $10^{-4}$, including leakage outside the computational subspace.

The objective of the proposed scheme is not to exceed the experimentally demonstrated fidelity of conventional microwave control. Rather, the results show that coherent errors intrinsic to discrete SFQ excitation and the weak anharmonicity of the transmon can be suppressed to a level comparable with high-fidelity microwave control while retaining the potential system-level advantages of cryogenic digital electronics, including reduced room-temperature hardware and more scalable control routing.

A key element of the approach is the use of combined sequences consisting of a regular train framed by two arbitrary ramps. This structure suppresses leakage while allowing only the ramps to be stored in the controller memory. Further memory reduction is obtained using repeating subsequences. A modular hardware implementation supporting both regular and arbitrary pulse generation is proposed, and its logical complexity is estimated from the JJ counts of the constituent SFQ cells.

The present results therefore establish that coherent leakage and control accuracy need not constitute a limiting factor of a memory-efficient SFQ control at the qubit level, providing a basis for further development of scalable cryogenic control systems for superconducting quantum processors.

Future work will focus on optimizing the controller architecture for reduced power consumption and chip area, as well as on clocking techniques to ensure the temporal stability of the control protocol. The results obtained pave the way for scalable control platforms for superconducting quantum processors.

\appendix
\section{Symmetry exploration}\label{app:symmetry}

The concept of symmetry, initially proposed in \citet{shillito}, is examined to determine whether the symmetric nature of DRAG can be exploited to identify more effective control sequences. Sequences obtained using the genetic algorithm are evaluated using a symmetry score defined by comparing the left and right ramps and the gates they implement. Because any single-qubit operation can be decomposed into a sequence of X- and Y-operations, only these operations are considered. The symmetry score measures the extent to which operations performed by the left ramp are compensated by those of the right ramp. For example, if the left ramp implements $X(-\pi/2)$, the right ramp should implement $X(\pi/2)$ to “close” the sequence and preserve symmetry. To quantify the number of operations that remain uncompensated in the combined sequence, the following symmetry-scoring algorithm is introduced.

\begin{enumerate}
    \item Decompose the ramp sequences into X- and Y-operations.
    \item Assign a weight $w^{X,Y}_i$ to each operation $i$: $\pi$-operations have a weight of 1, $\pi/2$-operations have a weight of 0.5, and $\pi/4$-operations have a weight of 0.25. X- and Y-operations are weighted independently.
    \item Compute the score for each ramp as the sum of all individual operation weights: $W^X=\sum_i w^X_i, \; W^Y=\sum_i w^Y_i$. For both ramps (left, LW and right, RW) calculate the score: $LW^X, LW^Y, RW^X, RW^Y$.
    \item Calculate the symmetry scores for the entire sequence: $S^X = |LW^X-RW^X|, \;S^Y = |LW^Y-RW^Y|, \;S^\Sigma = S^X+S^Y$.
\end{enumerate}

\begin{figure*}[t]
    \centering
    \includegraphics[width=0.9\linewidth]{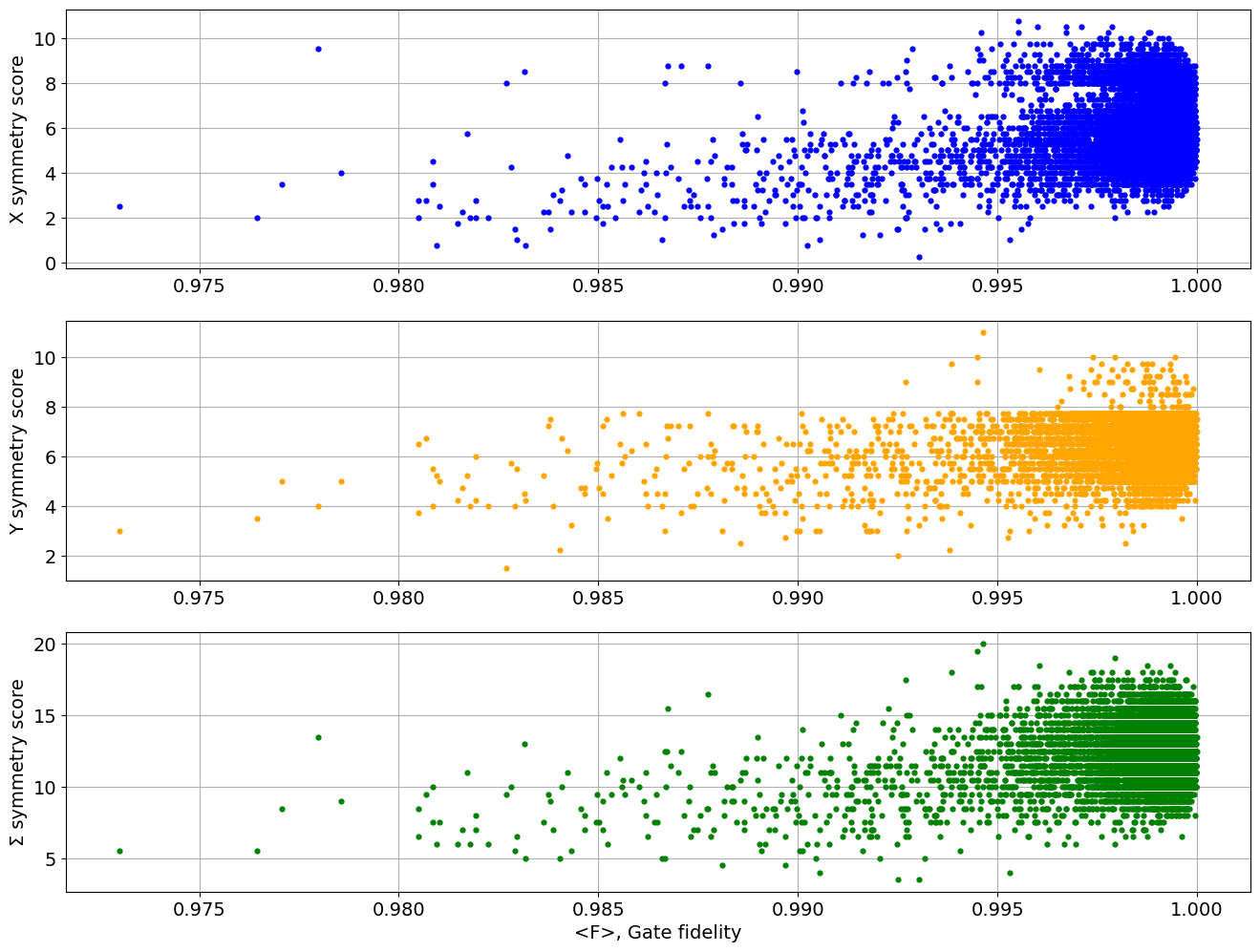}
    \caption{Symmetry exploration of the combined sequences found by the genetic algorithm.}
    \label{fig:9}
\end{figure*}

We have simulated the optimal search for combined sequences using the following set of parameters: main qubit frequency $\omega_{01}/2\pi = 5$ GHz, anharmonicity $\alpha/2\pi = 0.25$ GHz, single-pulse rotation angle $\Theta = 0.024$, generator frequency $\omega_g/2\pi = 40$ GHz, pulse width $\tau = 4$ ps, number of pulses in the regular part $M = 60$, and number of pulses in the ramp $N = 80$. All sequences with fidelity greater than 0.97 were analyzed for their symmetry score. The results of this analysis are presented in Fig.~\ref{fig:9}.

The search revealed almost no correlation between the symmetry score and gate fidelity across all metrics. It is possible to find sequences with high fidelity that exhibit both high and low symmetry scores. This result reinforces the idea that imposing symmetry constraints is not necessary for effective combined sequences.

\section{Parameter robustness study}\label{app:robustness}

To evaluate the robustness of the optimized SFQ control sequence, we performed an additional sensitivity analysis with respect to variations in the qubit frequency $\omega_{01}$, anharmonicity $\alpha$, single-pulse rotation angle $\Theta$, and pulse-timing jitter. In all cases, the highest fidelity sequence for the nominal parameter set \#2 of Table~\ref{tab:performance} was used, and no reoptimization was performed. The nominal parameter set was
$$
\omega_{01}/2\pi = 5~\mathrm{GHz}, \qquad
\alpha/2\pi = 0.25~\mathrm{GHz}, \qquad
\Theta = 0.030,
$$
with generator frequency $\omega_g/2\pi = 40~\mathrm{GHz}$ and
pulse width $\tau = 4~\mathrm{ps}$.

\begin{figure*}[t]
    \centering
    \includegraphics[width=0.9\linewidth]{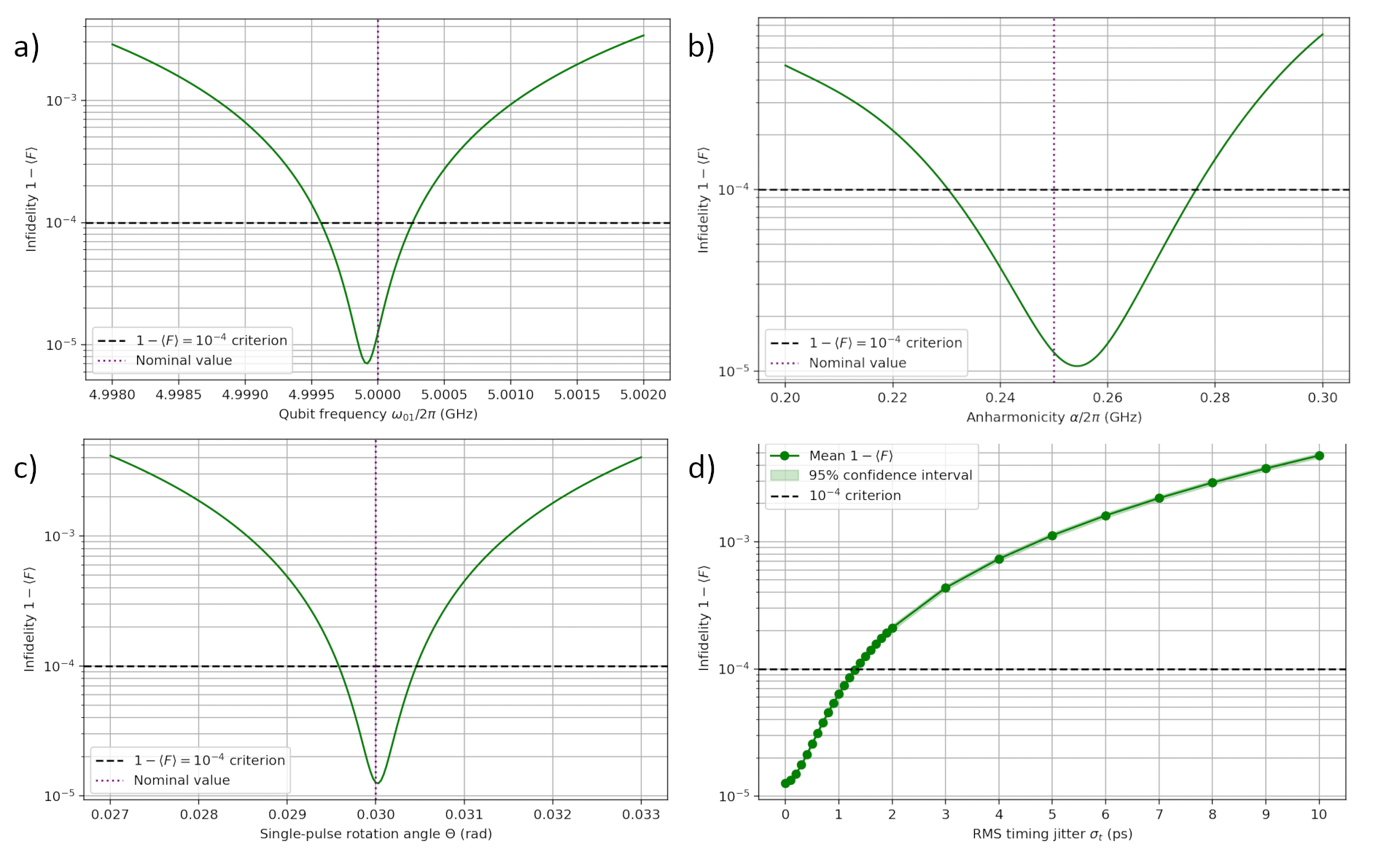}
    \caption{
    Robustness of the optimized $Y_{\pi/2}$ SFQ sequence.
    (a) Infidelity versus qubit frequency $\omega_{01}/2\pi$.
    (b) Infidelity versus anharmonicity $\alpha/2\pi$.
    (c) Infidelity versus single-pulse rotation angle $\Theta$.
    (d) Mean infidelity versus rms pulse-timing jitter $\sigma_t$,
    averaged over 1000 Monte Carlo realizations; the shaded region
    shows the 95\% confidence interval of the mean.     The horizontal dashed line marks $1-\braket{F}=10^{-4}$.
    }
    \label{fig:10}
\end{figure*}

The sequence itself is:
\begin{equation*}
seq = left \,\Vert\, (\mathtt{10000000})^{45} \,\Vert\, right,
\end{equation*}
\begin{align*}
left ={}& \mathtt{10010111\,01010011\,00000010\,11101100}
\nonumber\\
    & \mathtt{11101000\,11110111\,00011010},
\\[0.5em]
right ={}& \mathtt{10000011\,10010010\,10110111\,01010111}
\nonumber\\
    & \mathtt{10110010\,10001010\,01010001}.
\end{align*}
Figure~\ref{fig:10} summarizes the dependence of the infidelity $1-\braket{F}$ on each perturbation.

In Fig.~\ref{fig:10}(a), the qubit frequency was varied while all other parameters were kept fixed. The generator frequency was not retuned, so this panel characterizes sensitivity to qubit frequency detuning with respect to the nominal $40~\mathrm{GHz}$ clock. The fidelity criterion $ 1-\braket{F} < 10^{-4}$ is satisfied for
$$
4.99958~\mathrm{GHz}
\le
\omega_{01}/2\pi
\le
5.00026~\mathrm{GHz},
$$
corresponding to an allowed frequency offset of approximately $-0.42$ to $+0.26~\mathrm{MHz}$ around the nominal value.

In Fig.~\ref{fig:10}(b), the anharmonicity was varied while the remaining parameters were fixed. The same fidelity criterion is satisfied in the interval
$$
0.2305~\mathrm{GHz}
\le
\alpha/2\pi
\le
0.2760~\mathrm{GHz}.
$$
Thus, the optimized sequence remains below the $10^{-4}$ infidelity threshold over a relatively broad range of anharmonicities around the nominal value $\alpha/2\pi = 0.25~\mathrm{GHz}$.

In Fig.~\ref{fig:10}(c), the single-pulse rotation angle $\Theta$ was varied while all qubit parameters were kept fixed. The criterion
$1-\braket{F} < 10^{-4}$ is satisfied for
$$
0.02961 \le \Theta \le 0.03045.
$$
This corresponds to a tolerance of approximately $-1.3\%$ to $+1.5\%$ relative to the nominal value $\Theta = 0.030$.

Finally, Fig.~\ref{fig:10}(d) shows the sensitivity to pulse-timing jitter. Each nonzero SFQ pulse was independently shifted from its nominal arrival time according to
\begin{equation}
t_i = t_i^{(0)} + \delta t_i,
\qquad
\delta t_i \sim \mathcal{N}(0,\sigma_t^2),
\label{eq:jitter_model}
\end{equation}
where $\sigma_t$ is the rms timing jitter. The pulse width and pulse area were kept unchanged. The nominal digital clock grid and total sequence duration were kept fixed. Empty clock slots were unchanged, and the displacement of one pulse did not alter the nominal arrival times of subsequent pulses. For each value of $\sigma_t$, the sequence was simulated for 1000 independent Monte Carlo realizations, and the resulting infidelity was averaged. The shaded region in Fig.~\ref{fig:10}(d), corresponds to the 95\% confidence interval of the Monte Carlo mean.

The mean infidelity remains below $10^{-4}$ up to approximately $\sigma_t \approx 1.32~\mathrm{ps}$, while at least 95\% of the Monte Carlo realizations satisfy the criterion for $\sigma_t \le 0.7~\mathrm{ps}.$ As expected, the effect of timing jitter becomes significant once the jitter approaches the picosecond scale, i.e., a non-negligible fraction of the pulse duration and clock period.

Overall, the robustness study shows that the optimized sequence is most sensitive to timing and qubit-frequency errors, while it is substantially more tolerant to variations in anharmonicity and single-pulse rotation angle. These results further support the practical feasibility of the combined-sequence approach, while also identifying the timing stability of the SFQ generator as a key requirement for experimental implementation.

\section*{Acknowledgments}
The development of the main concept for qubit control was supported by the Russian Science Foundation, Russia, Grant No. 24-19-00187. This research partially used the computational resources of the HPC SEC shared research facilities at UNN. The development of controller schematics was carried out with the support of the Roadmap for Quantum computing (Contract No. 868-1.3-15/15-2021 dated October 5).

\section*{Conflict of Interest}
The authors have no conflicts to disclose.

\section*{Data Availability}
The data that support the findings of this study are available from the corresponding author upon reasonable request.

\bibliography{bibliography_reviewed}

\end{document}